\documentclass[12pt,leqno,a4paper]{article}
\usepackage{amsmath,amsthm,enumerate}
\usepackage[latin1]{inputenc}
\usepackage[T1]{fontenc}
\usepackage[english]{babel}

\newtheorem{theorem}{Theorem}[section]
\newtheorem{proposition}[theorem]{Proposition}
\newtheorem{lemma}[theorem]{Lemma}
\newtheorem{corollary}[theorem]{Corollary}
\newtheorem{definition}[theorem]{Definition}

\newtheorem{remark}[theorem]{Remark}


\newcommand*\proofnamestyle{\itshape}

\DeclareMathOperator{\tr}{Tr}

\DeclareMathOperator{\Var}{Var}
\DeclareMathOperator{\Corr}{Corr}

\begin{document}

    \title{Metric adjusted skew information}
      \author{Frank Hansen}
      \date{July 22, 2006\\
      {\tiny (Revised February 20, 2008)}}

      \maketitle

      \begin{abstract}
      We extend the concept of Wigner-Yanase-Dyson skew information to something
      we call ``metric adjusted skew information'' (of a state with respect to a
      conserved observable). This ``skew
      information'' is intended to be a non-negative quantity bounded by the variance
      (of an observable in a state) that vanishes for observables commuting with
      the state. We show that the skew information is a convex function on the
      manifold of states. It also satisfies other requirements, proposed by Wigner
      and Yanase, for an effective measure-of-information content of a state relative
      to a conserved observable. We establish a connection between the geometrical
      formulation of quantum statistics as proposed by Chentsov and Morozova and
      measures of quantum information as introduced by Wigner and Yanase and extended
      in this article. We show that the set of normalized Morozova-Chentsov functions
      describing the possible quantum statistics is a Bauer simplex and determine its
      extreme points. We determine a particularly simple skew information, the
      ``$ \lambda $-skew information,'' parametrized by a $ \lambda\in (0,1], $ and
      show that the convex cone this family generates coincides with the set of all
      metric adjusted skew informations.
      \\[1ex] Key words: Skew information, convexity, monotone
      metric, Morozova-Chentsov function, $ \lambda $-skew information.
      \end{abstract}

    \section{Introduction}

    In the mathematical model for a quantum mechanical system, the physical observables are
    represented by self-adjoint operators on a Hilbert space. The ``states'' (that is, the
    ``expectation functionals'' associated with the states) of the physical system are often
    ``modeled'' by the unit vectors in the underlying Hilbert space. So, if $ A $ represents
    an observable and $ x\in H $ corresponds to a state of the system, the expectation of
    $ A $ in that state is $ (Ax\mid x). $ For what we shall be proving, it will suffice
    to assume that our Hilbert space is finite dimensional and that the observables are
    self-adjoint operators, or the matrices that represent them, on that finite dimensional
    space. In this case, the states can be realized with the aid of the trace (functional)
    on matrices and an associated ``density matrix''. We denote by $ \tr(B) $ the usual
    trace of a matrix $ B $ (that is, $ \tr(B) $ is the sum of the diagonal entries of $ B). $
    The expectation functional of a state can be expressed as $ \tr(\rho A), $ where $ \rho $
    is a matrix, the density matrix associated with the state, and ``$ \tr(\rho A) $'' is the
    trace of the product $ \rho A $ of the two matrices $ \rho $ and $ A. $ (Henceforth, we
    write ``$ \tr\rho A $'' omitting the parentheses when they are clearly understood.)

    In \cite{kn:wigner:1952}, Wigner noticed that the obtainable accuracy of the measurement
    of a physical observable represented by an operator that does not commute with a conserved
    quantity (observable) is limited by the ``extent'' of that non-commutativity.
    Wigner proved it in the simple case
    where the physical observable is the $ x $-component of the spin of a
    spin one-half particle and the $ z $-component of the angular momentum is
    conserved. Araki and Yanase \cite{kn:araki:1960} demonstrated that this
    is a general phenomenon and pointed out, following Wigner's example, that
    under fairly general conditions an approximate measurement may be carried
    out.

    Another difference is that observables that commute with a conserved
    additive quantity, like the energy, components of the linear or angular
    momenta, or the electrical charge, can be measured easily and accurately by
    microscopic apparatuses (the analysis is restricted to one conserved
    quantity), while other observables can be only approximately measured by
    a macroscopic apparatus large enough to superpose sufficiently many
    states with different quantum numbers of the conserved quantity.

    Wigner and Yanase \cite{kn:wigner:1963} proposed finding a measure of our
    knowledge of a difficult-to-measure observable with respect to a
    conserved quantity. The quantum mechanical entropy is a measure of our
    ignorance of the state of a system, and minus the entropy can therefore
    be considered as an expression of our knowledge of the system. This
    measure has many attractive properties but does not take into account the
    conserved quantity.  In  particular, Wigner and Yanase wanted a measure
    that vanishes when the observable commutes with the conserved quantity.
    It should therefore not measure the effect of mixing in the classical
    sense as long as the pure states taking part in the mixing commute with
    the conserved quantity.  Only transition probabilities of pure states
    ``lying askew'' (to borrow from the introduction of \cite{kn:wigner:1963})
    to the eigenvectors of the conserved quantity should give
    contributions to the proposed measure.

    Wigner and Yanase discussed a number of requirements that such a measure
    should satisfy in order to be meaningful and suggested, tentatively, the
    skew information defined by
    \[
    I(\rho,A)=-{\textstyle\frac{1}{2}} \tr \bigl([\rho^{1/2}, A]^2\bigr),
    \]
    where $ [C,D] $ is the usual ``bracket notation'' for operators or
    matrices: $ [C,D]=CD-DC, $
    as a measure of the information contained in a state $ \rho $
    with respect to a conserved observable $ A. $ It manifestly vanishes when
    $ \rho $ commutes with $ A, $ and it is homogeneous in $ \rho. $

    The requirements Wigner and Yanase discussed, all reflected properties considered attractive
    or even essential. Since information is lost when separated systems are
    united such a measure should be decreasing under the mixing of states,
    that is, be convex in $ \rho. $ The authors proved this for the skew
    information, but noted that other measures may enjoy the same properties;
    in particular, the expression
    \[
    -{\textstyle\frac{1}{2}}\tr[\rho^p,A][\rho^{1-p},A]\qquad 0<p<1
    \]
    proposed by Dyson. Convexity of this expression in $ \rho $ became the
    celebrated Wigner-Yanase-Dyson conjecture which was later proved by Lieb
    \cite{kn:lieb:1973:1}. (See also \cite{kn:hansen:2006:3} for a truly
    elementary proof.)

    The measure should also be additive with respect to the aggregation of
    isolated subsystems and, for an isolated system, independent of time.
    These requirements are discussed in more detail in section
    \ref{subsection: measures of quantum information}. They are easily seen
    to be satisfied by the skew information.

    In the process that is the opposite of mixing, the information content
    should decrease. This requirement comes from thermodynamics where it is
    satisfied for both classical and quantum mechanical systems. It reflects
    the loss of information about statistical correlations between two
    subsystems when they are only considered separately. Wigner and Yanase
    conjectured that the skew information also possesses this property.  They
    proved it when the state of the aggregated system is pure\footnote{We
    subsequently demonstrated~\cite{kn:hansen:2007:2} that the conjecture
    fails for general mixed states.}.

    The aim of this article is to connect the subject of measures of quantum
    information as laid out by Wigner and Yanase with the geometrical
    formulation of quantum statistics by Chentsov, Morozova and Petz.

    The Fisher information measures the statistical distinguishability of
    probability distributions.  Let $ {\cal P}_n=\{p=(p_1,\dots,p_n)\mid
    p_i>0 \} $ be the (open) probability simplex with tangent space $ T{\cal
    P}_n. $ The Fisher-Rao metric is then given by \[ M_p(u,v)=\sum_{i=1}^n
    \frac{u_i v_i}{p_i}\qquad u,v\in  T{\cal P}_n.  \] Note that $
    u=(u_1,\dots,u_n)\in T{\cal P}_n $ if and only if $ u_1+\dots+u_n=0, $
    but that the metric is well-defined also on $ \mathbf R^n. $ Chentsov
    proved that the Fisher-Rao metric is the unique Riemannian metric
    contracting under Markov morphisms \cite{kn:censov:1982}.

    Since Markov morphisms represent coarse graining or randomization, it
    means that the Fisher information is the only Riemannian metric
    possessing the attractive property that distinguishability of probability
    distributions becomes more difficult when they are observed through a
    noisy channel.

    Chentsov and Morozova extended the analysis to quantum mechanics by
    replacing Riemannian metrics defined on the tangent space of the simplex
    of probability distributions with positive definite sesquilinear
    (originally bilinear) forms $ K_\rho $ defined on the tangent space of a
    quantum system, where $ \rho $ is a positive definite state. Customarily,
    $ K_\rho $ is extended to all operators (matrices) supported by the
    underlying Hilbert space, cf. \cite{kn:petz:1996:2, kn:hansen:2006:2} for
    details. Noisy channels are in this setting represented by stochastic
    (completely positive and trace preserving) mappings $ T, $ and the
    contraction property by the monotonicity requirement
    \[
    K_{T(\rho)}(T(A),T(A))\le K_\rho(A,A)
    \]
    is imposed for every stochastic
    mapping $ T:M_n(\mathbf C)\to M_m(\mathbf C). $ Unlike the classical
    situation, it turned out that this requirement no longer uniquely
    determines the metric. By the combined efforts of Chentsov, Morozova and
    Petz it is established that the monotone metrics are given on the form
    \begin{equation}\label{Morozova-Chentsov function}
    K_\rho(A,B)=\tr A^* c(L_\rho,R_\rho) B,
    \end{equation}
    where $ c $ is a so called
    Morozova-Chentsov function and $ c(L_\rho,R_\rho) $ is the function taken
    in the pair of commuting left and right multiplication operators (denoted
    $ L_\rho $ and $ R_\rho $ respectively) by $ \rho. $ The
    Morozova-Chentsov function is of the form
    \[
    c(x,y)=\frac{1}{y f(x y^{-1})}\qquad x,y >0,
    \] where $ f $ is a positive operator monotone
    function defined in the positive half-axis satisfying the functional
    equation
    \begin{equation}\label{functional equation}
    f(t)=tf(t^{-1})\qquad t>0.
    \end{equation}
    The function
    \[
    f(t)=\frac{t+2\sqrt{t}+1}{4}\qquad t>0
    \]
    is clearly operator monotone
    and satisfies (\ref{functional equation}). The associated
    Morozova-Chentsov function \[
    c^{WY}(x,y)=\frac{4}{(\sqrt{x}+\sqrt{y})^2}\qquad x,y>0 \] therefore
    defines a monotone metric \[ K_\rho^{WY}(A,B)=\tr A^*
    c^{WY}(L_\rho,R_\rho) B, \] which we shall call the Wigner-Yanase metric.
    The starting point of our investigation is the observation by Gibilisco
    and Isola \cite{kn:gibilisco:2003} that
    \[
    I(\rho,A)={\textstyle\frac{1}{8}}\tr i[\rho, A] c^{WY}(L_\rho,R_\rho) i[\rho, A].
    \]
    There is thus a
    relationship between the Wigner-Yanase measure of quantum information and
    the geometrical theory of quantum statistics. It is the aim of the
    present article to explore this relationship in detail. The main result
    is that all well-behaved measures of quantum information - including
    the Wigner-Yanase-Dyson skew informations - are given in this way for a
    suitable subclass of monotone metrics.

    \subsection{Regular metrics}

    \begin{definition}[Regular metric]
    We say that a symmetric monotone metric \cite{kn:morozova:1990, kn:petz:1996}
    on the state space of a quantum system is regular, if the corresponding Morozova-Chentsov function c
    admits a strictly positive limit
    \[
    m(c)=\lim_{t\to 0} c(t,1)^{-1}.
    \]
    We call $ m(c) $ the metric constant.
    \end{definition}

    We also say, more informally, that a Morozova-Chentsov function $ c $ is regular if $ m(c)>0. $
    The function $ f(t)=c(t,1)^{-1} $ is positive and operator monotone on the positive half-line and
    may be extended to the closed positive half-line. Thus the metric constant $ m(c)=f(0). $

    \begin{definition}[metric adjusted skew information]\hskip 1em\\
    Let $ c $ be the Morozova-Chentsov function  of a regular metric. We introduce
    the metric adjusted skew information $ I^c_\rho(A) $ by setting
    \begin{equation}\label{formula: metric adjusted skew information}
    \begin{array}{rl}
    I^c_\rho(A)&=\frac{m(c)}{2}\displaystyle K_\rho^c(i[\rho, A], i[\rho, A])\\[2ex]
    &=\frac{m(c)}{2}\displaystyle\tr i[\rho, A] c(L_\rho,R_\rho)i[\rho, A]
    \end{array}
    \end{equation}
    for every $ \rho\in\mathcal M_n $ (the manifold of states) and every self-adjoint $ A\in M_n(\mathbf C). $
    \end{definition}

    Note that the metric adjusted skew information is proportional to the square of the metric length,
    as it is calculated by the symmetric monotone
    metric $ K_\rho^c $ with Morozova-Chentsov function $ c, $ of the commutator $ i[\rho, A], $ and
    that this commutator belongs to the
    tangent space of the state manifold $ \mathcal M_n. $ Metric adjusted skew information is thus a
    non-negative quantity. If we consider the WYD-metric with Morozova-Chentsov function
    \[
    c^{WYD}(x,y)=\frac{1}{p(1-p)}\cdot\frac{(x^p-y^p)(x^{1-p}-y^{1-p})}{(x-y)^2}
    \qquad 0<p<1,
    \]
    then the metric constant $ m(c^{WYD})=p(1-p) $ and the metric adjusted skew information
    \[
    \begin{array}{rl}
    I^{c^{WYD}}_\rho(A)&=\frac{p(1-p)}{2}\displaystyle\tr i[\rho, A] c^{WYD}(L_\rho,R_\rho) i[\rho,A]\\[2ex]
    &=-\frac{1}{2}\tr [\rho^p,A][\rho^{1-p},A]
    \end{array}
    \]
    becomes the Dyson generalization of the Wigner-Yanase skew information\footnote{Hasegawa and Petz proved in \cite{kn:petz:1996:1}
    that the function $ c^{WYD} $ is a Morozova-Chentsov function. They also proved that
    the Wigner-Yanase-Dyson skew information is proportional to the (corresponding) quantum Fisher information of the commutator
    $ i[\rho,A]. $}. The choice of the factor $ m(c) $ therefore
    works also for $ p\ne 1/2. $ It is in fact a quite general construction, and the metric constant
    is related to the
    topological properties of the metric adjusted skew information close to the border of the state manifold. But it is
    difficult to ascertain these properties directly, so we postpone further
    investigation until having established that $ I^c_\rho(A) $ is a convex function in $ \rho. $
    Since the commutator $ i[\rho,A]=i(L_\rho-R_\rho)A $ we may rewrite the metric adjusted skew information as
    \begin{equation}\label{representation of metric adjusted skew information}
    \begin{array}{rl}
    I^c_\rho(A)&=\frac{m(c)}{2}\displaystyle\tr A(i(L_\rho-R_\rho))^* c(L_\rho,R_\rho) i(L_\rho-R_\rho) A\\[2ex]
    &=\frac{m(c)}{2}\displaystyle\tr A\, \hat c(L_\rho,R_\rho)A,
    \end{array}
    \end{equation}
    where
    \begin{equation}
    \hat c(x,y)=(x-y)^2 c(x,y)\qquad x,y>0.
    \end{equation}

    Before we can address these questions in more detail, we have to study various characterizations
    of (symmetric) monotone metrics.

    \section{Characterizations of monotone metrics}

    \begin{theorem}\label{theorem: canonical representation for g}
    A positive operator monotone decreasing function $ g $ defined in the
    positive half-axis and satisfying the functional equation
    \begin{equation}\label{functional equation for g}
    g(t^{-1})=t\cdot g(t)
    \end{equation}
    has a canonical representation
    \begin{equation}\label{canonical representation for g}
    g(t)=\int_0^1\left(\frac{1}{t+\lambda}+\frac{1}{1+t\lambda}\right)d\mu(\lambda),
    \end{equation}
    where $ \mu $ is a finite Borel measure with support in $ [0,1]. $
    \end{theorem}

    \begin{proof}
    The function $ g $ is necessarily of the form
    \[
    g(t)=\beta+\int_0^\infty \frac{1}{t+\lambda}\, d\mu(\lambda),
    \]
    where $ \beta\ge 0 $ is a constant and $ \mu $ is a positive Borel measure such that the integrals
    $ \int(1+\lambda^2)^{-1} d\mu(\lambda) $ and $ \int \lambda(1+\lambda^2)^{-1}d\mu(\lambda) $ are finite,
    cf. \cite[Page 9]{kn:hansen:2006:1}.
    We denote by $ \tilde\mu $ the measure obtained from $ \mu $ by removing a possible atom in zero. Then, by making the transformation $ \lambda\to \lambda^{-1}, $ we may write
    \[
    \begin{array}{rl}
    g(t)&=\displaystyle\beta+\frac{\mu(0)}{t}+\int_0^\infty\frac{1}{t+\lambda}\,d\tilde\mu(\lambda)\\[3ex]
    &=\displaystyle\beta+\frac{\mu(0)}{t}+\int_0^\infty\frac{1}{t+\lambda^{-1}}\cdot\frac{1}{\lambda^2}\,d\tilde\mu(\lambda^{-1})\\[3ex]
    &=\displaystyle\beta+\frac{\mu(0)}{t}+\int_0^\infty\frac{1}{1+t\lambda}\,d\nu(\lambda),
    \end{array}
    \]
    where $ \nu $ is the Borel measure given by $ d\nu(\lambda)=\lambda^{-1}d\tilde\mu(\lambda^{-1}). $
    Since $ g $ satisfies the functional equation (\ref{functional equation for g}) we obtain
    \[
    \beta+\mu(0)t+\int_0^\infty\frac{1}{1+t^{-1}\lambda}\,d\nu(\lambda)=t\beta+\mu(0)+\int_0^\infty\frac{t}{t+\lambda}\,d\tilde\mu(\lambda).
    \]
    By letting $ t\to 0 $ and since $ \nu $ and $ \tilde\mu $ have no atoms in zero, we obtain $ \beta=\mu(0) $ and
    consequently
    \[
    \int_0^\infty\frac{1}{t+\lambda}\,d\nu(\lambda)=\int_0^\infty\frac{1}{t+\lambda}\,d\tilde\mu(\lambda)\qquad t>0.
    \]
    By analytic continuation we realize that both measures  $ \nu $ and $ \tilde\mu $
    appear as the representing measure of an analytic function with
    negative imaginary part in the complex upper half plane. They
    are therefore, by the representation theorem for this class of
    functions, necessarily identical. We finally obtain
    \[
    \begin{array}{rl}
    g(t)&\displaystyle=\beta+\frac{\beta}{t}+\int_0^\infty\frac{1}{t+\lambda}\,d\tilde\mu(\lambda)\\[3ex]
    &\displaystyle=\beta+\frac{\beta}{t}
    +\int_0^1\frac{1}{t+\lambda}\,d\tilde\mu(\lambda)+\int_0^1 \frac{1}{t+\lambda^{-1}}\cdot\frac{1}{\lambda^2}\,d\tilde\mu(\lambda^{-1})\\[3ex]
    &\displaystyle=\beta+\frac{\beta}{t}+\int_0^1\frac{1}{t+\lambda}\,d\tilde\mu(\lambda)+\int_0^1 \frac{1}{1+t\lambda}\,d\nu(\lambda)\\[3ex]
    &\displaystyle=\beta+\frac{\beta}{t}+\int_0^1\left(\frac{1}{t+\lambda}+\frac{1}{1+t\lambda}\right)d\tilde\mu(\lambda)\\[3ex]
    &\displaystyle=\int_0^1\left(\frac{1}{t+\lambda}+\frac{1}{1+t\lambda}\right)d\mu(\lambda).
    \end{array}
    \]
    The statement follows since every function of this form obviously is operator monotone
    decreasing and satisfy the functional equation (\ref{functional equation for g}).
    We also realize that the representing measure $ \mu $ is uniquely defined.
    \end{proof}

    \begin{remark}\label{remark: construction of the measure}\rm
    Inspection of the proof of Theorem~\ref{theorem: canonical representation for g} shows that the
    Pick function $ -g(x)=-c(x,1) $ has the canonical representation
    \[
    -g(x)=-g(0) + \int_{-\infty}^0 \frac{1}{\lambda - t}\, d\mu(-\lambda).
    \]
    The representing measure therefore appears as $ 1/\pi $ times the
    limit measure of the imaginary part of the analytic continuation $ -g(z) $ as $ z $ approaches the
    closed negative half-axis from above, cf. for example \cite{kn:donoghue:1974}. The measure $ \mu $
    in (\ref{canonical representation for g}) therefore appears as the image of the representing measure's restriction
    to the interval $ [-1, 0] $ under the transformation $ \lambda\to -\lambda. $
    \end{remark}

    We define, in the above setting, an equivalent Borel measure $ \mu_g $ on the closed interval $ [0,1] $ by setting
    \begin{equation}
    d\mu_g(\lambda)=\frac{2}{1+\lambda}\, d\mu(\lambda)
    \end{equation}
    and obtain:

    \begin{corollary}
    A positive operator monotone decreasing function $ g $ defined in the
    positive half-axis and satisfying the functional equation
    (\ref{functional equation for g}) has a canonical representation
    \begin{equation}\label{variant canonical representation for g}
    g(t)=\int_0^1 \frac{1+\lambda}{2}\left(\frac{1}{t+\lambda}+\frac{1}{1+t\lambda}\right)d\mu_g(\lambda),
    \end{equation}
    where $ \mu_g $ is a finite Borel measure with support in $ [0,1]. $ The function $ g $ is normalized
    in the sense that $ g(1)=1, $ if and only if $ \mu_g $ is a probability measure.
    \end{corollary}

    \begin{corollary}
    A Morozova-Chentsov function $ c $ allows a canonical
    representation of the form
    \begin{equation}\label{canonical representation for c}
    c(x,y)=\int_0^1 c_\lambda(x,y)\, d\mu_c(\lambda)\qquad x,y>0,
    \end{equation}
    where $ \mu_c $ is a finite Borel measure on $ [0,1] $ and
    \begin{equation}\label{extreme Morozova-Chentsov functions}
    c_\lambda(x,y)=\frac{1+\lambda}{2}\left(\frac{1}{x+\lambda y}+\frac{1}{\lambda x+y}\right)\qquad
    \lambda\in[0,1].
    \end{equation}
    The Morozova-Chentsov function $ c $ is normalized in the sense that $ c(1,1)=1 $ (corresponding to a Fisher
    adjusted metric), if and only if $ \mu_c $ is a probability measure.
    \end{corollary}

    \begin{proof}
    A Morozova-Chentsov function is of the form $ c(x,y)=y^{-1} f(xy^{-1})^{-1}, $
    where $ f $ is a positive operator monotone function defined in
    the positive half-axis and satisfying the functional equation $ f(t)=tf(t^{-1}). $ The function
    $ g(t)=f(t)^{-1} $ is therefore operator monotone
    decreasing and satisfies the functional equation (\ref{functional equation for g}).
    It is consequently of the form (\ref{variant canonical representation for g}) for some finite
    Borel measure $ \mu_g. $ Since also $ c(x,y)=y^{-1} g(xy^{-1}) $ the assertion follows
    by setting $ \mu_c=\mu_g. $
    \end{proof}

    We have shown that the set of normalized Morozova-Chentsov functions is a Bauer simplex,
    and that the extreme points exactly are the functions of the form (\ref{extreme Morozova-Chentsov functions}).

    \begin{theorem} We exhibit the measure $ \mu_c $ in the canonical representation
    (\ref{canonical representation for c}) for a number of Morozova-Chentsov functions.

    \begin{itemize}

    \item[1.]The Wigner-Yanase-Dyson metric with (normalized) Morozova-Chentsov function
    \[
    c(x,y)=\frac{1}{p(1-p)}\cdot\frac{(x^p-y^p)(x^{1-p}-y^{1-p})}{(x-y)^2}
    \]
    is represented by
    \[
    d\mu_c(\lambda)=\frac{2\sin p\pi}{\pi p (1-p)}\cdot\frac{\lambda^p + \lambda^{1-p}}{(1+\lambda)^3}\,d\lambda
    \]
    for $ 0<p<1. $

    The Wigner-Yanase metric is obtained by setting $ p=1/2 $ and it is represented by
    \[
    d\mu_c(\lambda)=\frac{16\lambda^{1/2}}{\pi (1+\lambda)^3}\, d\lambda.
    \]

    \item[2.] The Kubo metric with (normalized) Morozova-Chentsov function
    \[
    c(x,y)=\frac{\log x-\log y}{x-y}
    \]
    is represented by
    \[
    d\mu_c(\lambda)=\frac{2}{(1+\lambda)^2}\,d\lambda.
    \]

    \item[3.] The increasing bridge with (normalized) Morozova-Chentsov functions
    \[
    c_\gamma(x,y)=x^{-\gamma}y^{-\gamma}\left(\frac{x+y}{2}\right)^{2\gamma-1}
    \]
    is represented by
    \[
    \left\{\begin{array}{rll}
    \mu_c&=\delta(\lambda-1) &\gamma=0\\[1ex]
    d\mu_c(\lambda)&\displaystyle=\frac{2\sin\gamma\pi}{(1+\lambda)\pi} \lambda^{-\gamma}
    \left(\frac{1-\lambda}{2}\right)^{2\gamma-1}d\lambda\qquad &0<\gamma<1\\[2ex]
    \mu_c&\displaystyle=\delta(\lambda) &\gamma=1,
    \end{array}\right.
    \]
    where $ \delta $ is the Dirac measure with unit mass in zero.
    \end{itemize}
    \end{theorem}

    \begin{proof}We calculate the measures by
    the method outlined in Remark~\ref{remark: construction of the measure}.

    1. For the Wigner-Yanase-Dyson metric we therefore consider the analytic continuation
    \[
    -g(r e^{i\phi})=-c(r e^{i\phi}, 1)=
    \frac{-1}{p(1-p)}\cdot\frac{(r^p e^{ip\phi}-1)(r^{1-p} e^{i(1-p)\phi}-1)}{(r e^{i\phi}-1)^2}
    \]
    where $ r>0 $ and $ 0<\phi<\pi. $ We calculate the imaginary part and note that $ r\to-\lambda $
    and $ \phi\to\pi $ for
    $ z\to \lambda<0. $ We make sure that the representing measure has no atom in zero and obtain
    the desired expression by
    tedious but elementary calculations.

    2. For the Kubo metric we consider
    \[
    -g(x)=-c(x,1)=-\frac{\log x}{x-1}
    \]
    and calculate the imaginary part
    \[
    -\Im g(re^{i\phi})=\frac{2r\log r \sin\phi + \phi -\phi r \cos\phi}{r^2-2r\cos\phi +1}
    \]
    of the analytic continuation. It converges towards $ \pi/(1-\lambda) $ for $ z\to\lambda<0 $ and
    towards $ \pi/2 $
    for $ z=re^{i\pi}\to 0. $ The representing measure has therefore no atom in zero, and
     $ d\mu(\lambda)=d\lambda/(1+\lambda) $
    which may be verified by direct calculation.

    3. For the increasing bridge we consider
    \[
    -g_\gamma(x)=-c_\gamma(x,1)=-x^{-\gamma}\left(\frac{x+1}{2}\right)^{2\gamma-1}
    \]
    and calculate the imaginary part
    \[
    -\Im g_\gamma(re^{i\phi})=-r^{-\gamma}r_1^{2\gamma-1}\exp i(-\gamma\phi+(2\gamma-1)\theta)
    \]
    of the analytic continuation, where
    \[
    r_1={\textstyle\frac{1}{2}}(r^2+2r\cos\phi+1)^{1/2}\quad\text{and}\quad\theta=\arctan\frac{rsin\phi}{1+r\cos\phi}.
    \]
    We first note that $ \theta=\pi/2 $ and $ r_1=(r \sin\phi)/2 $ for $ \lambda=-1, $ and that $ \theta\to 0 $ and $ r_1\to (1+\lambda)/2 $
    for $ -1<\lambda\le 0. $ The statement now follows by examination of the different cases.
    \end{proof}

    In the reference \cite{kn:hansen:2006:2} we proved the
    following exponential representation of the Morozova-Chentsov
    functions.

    \begin{theorem}\label{theorem: set of Morozova-Chentsov functions}
    A Morozova-Chentsov function $ c $ admits a canonical representation
    \begin{equation}\label{canonical representation of c in terms of h}
    c(x,y)=\frac{C_0}{x+y}
    \exp\int_0^1\frac{1-\lambda^2}{\lambda^2+1}\cdot
    \frac{x^2+y^2}{(x+\lambda y)(\lambda x +y)}h(\lambda)\,d\lambda
    \end{equation}
    where $ h:[0,1]\to[0,1] $ is a measurable function and $ C_0 $ is a positive constant.
    Both $ C_0 $ and the equivalence class containing $ h $
    are uniquely determined by $ c. $ Any function $ c $
    on the given form is a Morozova-Chentsov function.
    \end{theorem}

    \begin{theorem} We exhibit the constant $ C_0 $ and the representing function $ h $ in the canonical representation
    (\ref{canonical representation of c in terms of h}) for a number of Morozova-Chentsov functions.

    \begin{itemize}

    \item[1.]The Wigner-Yanase-Dyson metric with Morozova-Chentsov function
    \[
    c(x,y)=\frac{1}{p(1-p)}\cdot\frac{(x^p-y^p)(x^{1-p}-y^{1-p})}{(x-y)^2}
    \]
    is represented by
    \[
    C_0=\frac{\sqrt{2}}{p(1-p)} \left(1-\cos p\frac{\pi}{2}\right)^{1/2} \left(1-\cos(1-p)\frac{\pi}{2}\right)^{1/2}
    \]
    and
     \[
    h(\lambda)=\frac{1}{\pi}\arctan\frac{(\lambda^p + \lambda^{1-p})\sin p\pi}{1-\lambda-(\lambda^p - \lambda^{1-p})\cos p\pi}\qquad 0<\lambda<1,
    \]
    for $ 0<p<1. $  Note that $ 0\le h\le 1/2. $

    The Wigner-Yanase metric is obtained by setting $ p=1/2 $ and is represented by
    \[
    C_0=4(\sqrt{2}-1)
    \]
    and
    \[
    h(\lambda)=\frac{1}{\pi}\arctan\frac{2\lambda^{1/2}}{1-\lambda}\qquad 0<\lambda<1.
    \]

    \item[2.] The Kubo metric with Morozova-Chentsov function
    \[
    c(x,y)=\frac{\log x-\log y}{x-y}
    \]
    is represented by
    \[
    C_0=\frac{\pi}{2}\quad\text{and}\quad
    h(\lambda)=\frac{1}{2}-\frac{1}{\pi}\arctan\left(-\frac{\log\lambda}{\pi}\right).
    \]
     Note that $ 0\le h\le 1/2. $

    \item[3.] The increasing bridge with Morozova-Chentsov functions
    \[
    c_\gamma(x,y)=x^{-\gamma}y^{-\gamma}\left(\frac{x+y}{2}\right)^{2j-1}
    \]
    is represented by
    \[
    C_0=2^{1-\gamma}\quad\text{and}\quad h(\lambda)=\gamma,\qquad  0\le\gamma\le 1.
    \]
    Setting $ \gamma=0, $ we obtain that the Bures metric with
    Morozova-Chentsov function $ c(x,y)=2/(x+y) $ is represented by
    $ C_0=2 $ and $ h(\lambda)=0. $

    \end{itemize}
    \end{theorem}

    \begin{proof}
    The analytic continuation of the operator monotone function $ g(x)=\log f(x) $ into the upper complex plane,
    where $ f(x)=c(x,1)^{-1} $ is the operator monotone
    function representing \cite{kn:petz:1996:2} the Morozova-Chentsov function, has bounded imaginary part.
    The representing measure of the Pick function $ g $ is therefore absolutely continuous with respect to
     Lebesgue measure. Since
    $ f $ satisfies the functional equation $ f(t)=t f(t^{-1}) $ we only need to consider the restriction of
     the measure to the interval $ [-1, 0], $
    and the function $ h $ appears \cite{kn:hansen:2006:2} as the image under the transformation
    $ \lambda\to -\lambda $ of the Radon-Nikodym derivative.
    In the same reference it is shown that the constant $ C_0=\sqrt{2} e^{-\beta} $ where $ \beta=\Re\log f(i). $

    1. For the Wigner-Yanase-Dyson metric the corresponding operator monotone function
    \[
    f(x)=\frac{1}{c(x,1)}=p(1-p)\frac{(x-1)^2}{(x^p-1)(x^{1-p}-1)}
    \]
    and we calculate by tedious but elementary calculations
    \[
    \lim_{z\to\lambda}\Im\log f(z)=-\frac{1}{2i}\log H\qquad \lambda\in(-1,0),
    \]
    where
    \[
    H=\frac{N}
    {((-\lambda)^{2p}-2(-\lambda)^p\cos p\pi+1)((-\lambda)^{2(1-p)}-2(-\lambda)^{(1-p)}\cos(1-p)\pi+1)}
    \]
    and
    \[
    \begin{array}{rl}
    N&=(-\lambda)^2 + 2(-\lambda)^{1+p} e^{ip\pi}+(-\lambda)^{2p} e^{2ip\pi}-2(-\lambda)^{2-p} e^{-ip\pi}\\[1ex]
    &+\,4\lambda - 2(-\lambda)^p e^{ip\pi}+(-\lambda)^{2(1-p)} e^{-2ip\pi} + 2(-\lambda)^{1-p} e^{-ip\pi} + 1
    \end{array}
    \]
    happens to be the square of the complex number
    \[
    (1+\lambda) - ((-\lambda)^p - (-\lambda)^{1-p})\cos p\pi  -i((-\lambda)^p + (-\lambda)^{1-p})\sin p\pi
    \]
    with positive real part and negative imaginary part.
    Since $ H $ has modulus one we can therefore write
    \[
    H=e^{-2i\theta}\qquad \lambda\in(-1,0),
    \]
    where $ 0<\theta<\pi/2 $ and
    \[
    \tan\theta=\frac{((-\lambda)^p + (-\lambda)^{1-p})\sin p\pi}{1+\lambda-((-\lambda)^p - (-\lambda)^{1-p})\cos p\pi}
    \]
    which implies the expression for $ h. $ The constant $ C_0 $ is obtained
    by a simple calculation.

    2. For the Kubo metric the corresponding operator monotone function
    \[
    f(x)=\frac{1}{c(x,1)}=\frac{x-1}{\log x}
    \]
    and we obtain by setting $ z=r e^{i\phi} $ and $ z-1=r_1 e^{i\phi_1} $ the expression
    \[
    \Im\log f(z)=\frac{1}{2i}\left(\log\frac{\log r - i\phi}{\log r +i\phi} +2i\phi_1\right)\qquad
    0<\phi<\phi_1<\pi.
    \]
    Since
    \[
    \log\frac{\log r - i\phi}{\log r +i\phi}\to \log\frac{\log(-\lambda) - i\pi}{\log(-\lambda) +i\pi}
    \]
    for $ z\to\lambda\in(-1,0) $ and
    \[
    \frac{\log(-\lambda) - i\pi}{\log(-\lambda) +i\pi}=e^{-2i\theta}\quad\text{where}\quad\tan\theta=\frac{\pi}{\log(-\lambda)}
    \]
    we obtain
    \[
    \lim_{z\to\lambda}\Im\log f(z)=\pi-\theta\qquad\frac{\pi}{2}<\theta<\pi,
    \]
    and thus
    \[
    h(\lambda)=1-\frac{1}{\pi}\arctan\frac{\pi}{\log\lambda}.
    \]
    The constant $ C_0 $ is obtained by a straightforward calculation.

    3. The statement for the increasing bridge was proved in \cite{kn:hansen:2006:2}.
    \end{proof}

    \section{Convexity statements}

    \begin{proposition}
    Every Morozova-Chentsov function $ c $ is operator convex, and the mappings
     \[
    (\rho,\delta)\to \tr A^* c(L_\rho,R_\delta) A
    \]
    and
    \[
    \rho\to K_\rho^c(A,A)
    \]
    defined on the state manifold are convex for arbitrary $ A\in M_n(\mathbf C). $
    \end{proposition}

    \begin{proof}
    Let $ c $ be a Morozova-Chentsov function. Since inversion is operator convex, it follows from the
    representation given in (\ref{canonical representation for c}) that $ c $ as a function of two variables is operator convex.
    The two assertions now follows from \cite[Theorem 1.1]{kn:hansen:2006:3}.
    \end{proof}

    \begin{lemma}\label{lemma: operator convex functions}
    Let $ \lambda\ge 0 $ be a constant. The functions of two variables
    \[
    f(t,s)=\frac{t^2}{t+\lambda s}\quad\mbox{and}\quad
    g(t,s)=\frac{ts}{t+\lambda s}
    \]
    are operator convex respectively operator concave on $ (0,\infty)\times(0,\infty). $
    \end{lemma}

    \begin{proof}
    The first statement is an application of the convexity, due to Lieb and Ruskai, of the mapping $ (A,B)\to
    AB^{-1}A. $ Indeed, setting
    \[
    C_1=A_1\otimes I_2+\lambda I_1\otimes B_1\quad\mbox{and}\quad C_2=A_2\otimes I_2+\lambda I_1\otimes B_2
    \]
    we obtain
    \[
    \begin{array}{l}
    f(t A_1+(1-t)A_2, t B_1+(1-t)B_2)\\[2ex]
    =\displaystyle \Bigl((t A_1+(1-t)A_2)\otimes I_2\Bigr)
     (t C_1+(1-t) C_2 )^{-1} \Bigl((t A_1+(1-t)A_2)\otimes I_2\Bigr)\\[2ex]
     \le t (A_1\otimes I_2) C_1^{-1} (A_1\otimes I_2) + (1-t)(A_2\otimes I_2) C_2^{-1} (A_2\otimes I_2)\\[2ex]
     =t f(A_1,B_1)+(1-t)f(A_2,B_2)\qquad t\in[0,1].
    \end{array}
    \]
    The second statement is a consequence of the concavity of the harmonic mean
    \[
    H(A,B)=2(A^{-1}+B^{-1})^{-1}.
    \]
    Indeed, we may assume $ \lambda>0 $ and obtain
    \[
    \begin{array}{l}
     g(t A_1+(1-t)A_2, t B_1+(1-t)B_2)\\[1.5ex]
     \displaystyle=\frac{1}{2} H\Bigl(t(\lambda^{-1}A_1\otimes I_2)+
     (1-t)(\lambda^{-1}A_2\otimes I_2), t(I_1\otimes B_1)+(1-t)(I_1\otimes B_2)\Bigr)\\[2ex]
     \displaystyle\ge t\frac{1}{2} H(\lambda^{-1}A_1\otimes I_2,I_1\otimes B_1)+
     (1-t)\frac{1}{2} H(\lambda^{-1}A_2\otimes I_2,I_1\otimes B_2)\\[2ex]
     =t g(A_1,B_1)+(1-t)g(A_2,B_2)
    \end{array}
    \]
    for $ t\in(0,1]. $
    \end{proof}

    \begin{proposition}\label{c hat is operator convex}
    Let  $ c $ be a Morozova-Chentsov function. The function of two
    variables
    \[
    \hat c(x,y)=(x-y)^2 c(x,y)\qquad x,y>0
    \]
    is operator convex.
    \end{proposition}

    \begin{proof}
    A Morozova-Chentsov function $ c $ allows the representation
    (\ref{canonical representation for c}) where $ \mu $ is some
    finite Borel measure with support in $ [0,1]. $ Since
    \[
    \frac{(x-y)^2}{x+\lambda y}=\frac{x^2+y^2-2xy}{x+\lambda y}
    \]
     by Lemma \ref{lemma: operator convex functions} is a sum of operator convex functions
    the assertion follows.
    \end{proof}

    \begin{proposition}\label{proposition: decomposition of c}
    Let $ c $ be a regular Morozova-Chentson function. We may write $ \hat c(x,y)=(x-y)^2 c(x,y) $ on the form
    \begin{equation}\label{formula: decomposition for c hat}
    \hat c(x,y)=\frac{x+y}{m(c)}-d_c(x,y)\qquad x,y>0,
    \end{equation}
    where the positive symmetric function
    \begin{equation}\label{the function d representing a regular metric}
    d_c(x,y)=\int_0^1 xy\cdot c_\lambda(x,y)\frac{(1+\lambda)^2}{\lambda}\,d\mu_c(\lambda)
    \end{equation}
    is operator concave in the first quadrant, and the finite Borel measure $ \mu_c $ is the representing
    measure in (\ref{canonical representation for c}) of the Morozova-Chentsov function $ c. $
    In addition, we obtain the expression
    \begin{equation}\label{skew information in terms of d}
    \begin{array}{rl}
    I^c_\rho(A)&=\frac{m(c)}{2}\displaystyle\tr A\hat c(L_\rho,R_\rho)A\\[2ex]
    &=\tr\rho A^2-\frac{m(c)}{2}\displaystyle\tr A\, d_c(L_\rho,R_\rho)A
    \end{array}
    \end{equation}
    for the metric adjusted skew information.
    \end{proposition}

    \begin{proof}
    We first notice that
    \begin{equation}\label{integral formula for the metric constant}
    \int_0^1 \frac{(1+\lambda)^2}{2\lambda}\,d\mu_c(\lambda)=\lim_{t\to 0} c(t,1)=\frac{1}{m(c)}
    \end{equation}
    and obtain
    \[
    \begin{array}{rl}
    d_c(x,y)&\displaystyle=\frac{x+y}{m(c)}-\hat c(x,y)\\[3ex]
    &\displaystyle=\frac{x+y}{m(c)}-(x-y)^2 c(x,y)\\[3ex]
    &\displaystyle=(x+y)\int_0^1\frac{(1+\lambda)^2}{2\lambda}\,d\mu_c(\lambda)
    -(x-y)^2\int_0^1 c_\lambda(x,y)\,d\mu_c(\lambda)\\[3ex]
    &\displaystyle=\int_0^1\left((x+y)\frac{(1+\lambda)^2}{2\lambda}-(x-y)^2 c_\lambda(x,y)\right)\,d\mu_c(\lambda).
    \end{array}
    \]
    The asserted expression of $ d_c $ then follows by a simple calculation and the definition of
    $ c_\lambda(x,y) $ as given in (\ref{extreme Morozova-Chentsov functions}). The function $ d_c $ is operator concave
    in the first quadrant by Proposition~\ref{c hat is operator convex}.
    \end{proof}

    \begin{definition}
    We call the function $ d_c $ defined in (\ref{the function d representing a regular metric}) the
    representing function for the metric adjusted skew information $ I^c_\rho(A) $ with (regular)
    Morozova-Chentsov function $ c. $
    \end{definition}

    We introduce for $ 0<\lambda\le 1 $ the $ \lambda $-skew information $ I_\lambda(\rho, A) $ by setting
    \[
    I_\lambda(\rho, A)=I^{c_\lambda}_\rho(A).
    \]
    The metric is regular with metric constant $ m(c_\lambda)=2\lambda(1+\lambda)^{-2} $ and the representing
    measure $ \mu_{c_\lambda} $ is the Dirac measure in $ \lambda. $ The representing function for the
    metric adjusted skew information is thus given by
    \[
    d_{c_\lambda}(x,y)=xy\cdot c_\lambda(x,y) \frac{(1+\lambda)^2}{\lambda}
    =\frac{m(c_\lambda)}{2}\, xy\cdot c_\lambda(x,y).
    \]
    If we set
    \begin{equation}
    f_\lambda(x,y)=xy\cdot c_\lambda(x,y)=\frac{1+\lambda}{2}\left(\frac{xy}{x+\lambda y}+\frac{xy}{\lambda x+y}\right)
    \qquad x,y>0,
    \end{equation}
    we therefore obtain the expression
    \begin{equation}
    I_\lambda(\rho, A)=\tr\rho A^2 - \tr A f_\lambda(L_\rho, R_\rho) A
    \end{equation}
    for the $ \lambda $-skew information.

    \begin{corollary} Let $ c $ be a regular Morozova-Chentsov function. The metric adjusted
    skew information may be written on the form
    \[
    I^c_\rho(A)={\textstyle\frac{m(c)}{2}}\int_0^1 I_\lambda(\rho, A) \frac{(1+\lambda)^2}{\lambda}\, d\mu_c(\lambda),
    \]
    where $ \mu_c $ is the representing measure and $ m(c) $ is the metric constant.
    \end{corollary}

    \begin{proof} By applying the expressions in (\ref{skew information in terms of d}) and
    (\ref{the function d representing a regular metric}) together with the observation in
    (\ref{integral formula for the metric constant}) we obtain
    \[
    \begin{array}{rl}
    I^c_\rho(A)&=\displaystyle\tr\rho A^2 -{\textstyle\frac{m(c)}{2}}\int_0^1 \tr A f_\lambda(L_\rho,R_\rho)
    A\frac{(1+\lambda)^2}{\lambda}\, d\mu_c(\lambda)\\[2ex]
    &=\frac{m(c)}{2}\displaystyle (\tr\rho A^2 - \tr Af_\lambda(L_\rho,R_\rho)A)
    \frac{(1+\lambda)^2}{\lambda}\, d\mu_c(\lambda)
    \end{array}
    \]
    and the assertion follows.
    \end{proof}

    \subsection{Measures of quantum information}\label{subsection: measures of quantum information}

    The next result is a direct generalization of the Wigner-Yanase-Dyson-Lieb convexity theorem.
    \begin{theorem} Let $ c $ be a regular Morozova-Chentsov function.
    The metric adjusted skew information is a convex function,
    $ \rho\to I^c_\rho(A), $ on the manifold of states for any self-adjoint $ A\in M_n(\mathbf C). $
    \end{theorem}

    \begin{proof}
    The function $ \hat c(x,y)=(x-y)^2 c(x,y) $ is  by Proposition \ref{c hat is operator convex} operator convex.
    Applying the representation of the metric adjusted skew information given
    in (\ref{representation of metric adjusted skew information}), the assertion now
    follows from \cite[Theorem 1.1]{kn:hansen:2006:3}.
    \end{proof}

    The above proof is particularly transparent for the Wigner-Yanase-Dyson
    metric, since the function
    \[
    \begin{array}{rl}
    \hat
    c^{WYD}(\lambda,\mu)&\displaystyle=\frac{1}{p(1-p)}(\lambda^p-\mu^p)(\lambda^{1-p}-\mu^{1-p})\\[2ex]
    &\displaystyle=\frac{1}{p(1-p)}(2-\lambda^p\mu^{1-p}-\lambda^{1-p}\mu^p)
    \end{array}
    \]
    is operator convex by the simple argument given in \cite[Corollary 2.2]{kn:hansen:2006:3}.

    Wigner and Yanase \cite{kn:wigner:1963} discussed a number of other conditions
    which a good measure of the quantum information contained in a state with respect to
    a conserved observable should satisfy, but noted that convexity was the most obvious
    but also the most restrictive and difficult condition. In
    addition to the convexity requirement an information measure should be additive with respect
    to the aggregation of isolated systems.  Since the state of the aggregated system
    is represented by $ \rho=\rho_1\otimes\rho_2 $ where $ \rho_1 $ and $ \rho_2 $ are the states
    of the systems to be united, and the conserved quantity
    $ A=A_1\otimes 1 + 1\otimes A_2 $ is additive in its components, we obtain
    \[
    [\rho,A]=[\rho_1, A_1]\otimes \rho_2 + \rho_1\otimes [\rho_2, A_2].
    \]
    Inserting $ \rho $ and $ A, $ as above, in the definition of the metric adjusted skew information
    (\ref{formula: metric adjusted skew information}), we obtain
    \[
    \begin{array}{rl}
    I^c_\rho(A)&=\frac{m(c)}{2}\displaystyle\tr\Bigl(i[\rho_1, A_1]\otimes \rho_2 +
    \rho_1\otimes i[\rho_2, A_2]\Bigr)\\[1.5ex]
    &\hskip 3em c(L_{\rho_1},R_{\rho_1})\otimes c(L_{\rho_2},R_{\rho_2})\Bigl(i[\rho_1, A_1]\otimes \rho_2 +
    \rho_1\otimes i[\rho_2, A_2]\Bigr).
    \end{array}
    \]
    The cross terms vanish because of the cyclicity  of the trace,
    and since $ \rho_1 $ and $ \rho_2 $ have unit trace we obtain
    \[
    I^c_\rho(A)=I^c_{\rho_1}(A_1)+I^c_{\rho_2}(A_2)
    \]
    as desired. The metric adjusted skew information for an isolated system should also be independent of time.
    But a conserved quantity $ A $ in an isolated system commutes with the Hamiltonian $ H, $
    and since the time evolution of $ \rho $ is given by $ \rho_t=e^{itH}\rho e^{-itH} $ we readily obtain
    \[
    I^c_{\rho_t}(A)=I^c_\rho(A)\qquad t\ge 0
    \]
    by using the unitary invariance of the metric adjusted skew information.

    The variance $ \Var_\rho(A) $ of a conserved observable $ A $ with respect to a state $ \rho $ is defined by setting
    \[
    \Var_\rho(A)=\tr\rho A^2 - (\tr\rho A)^2.
    \]
    It is a concave function in $ \rho. $

    \begin{theorem}\label{theorem: the skew information is bounded by the variance}
    Let $ c $ be a regular Morozova-Chentsov function. The metric adjusted skew information $ I^c(\rho,A) $
    may for each conserved (self-adjoint) variable $ A $  be extended from the state manifold
    to the state space. Furthermore,
    \[
    I^c_\rho(A)=\Var_\rho(A)
    \]
    if $ \rho $ is a pure state, and
    \[
    0\le I^c_\rho(A)\le\Var_\rho(A)
    \]
    for any density matrix $ \rho. $
    \end{theorem}

    \begin{proof}
    We note that the representing function $ d $ in (\ref{the function d representing a regular metric})
    may be extended to a continuous operator concave function
    defined in the closed first quadrant with $ d(t,0)=d(0,t)=0 $ for every $ t\ge 0, $ and that $ d(1,1)=2/m(c). $
    Since a pure state is a one-dimensional projection $ P, $
    it follows from the representation in (\ref{representation of metric adjusted skew information})
    and the formula (\ref{formula: decomposition for c hat}) that
    \[
    \begin{array}{rl}
    I^c_P(A)&=\frac{m(c)}{2}\displaystyle\tr\left(\frac{APA+AAP}{m(c)}-d(1,1) APAP\right)\\[2ex]
    &\displaystyle=\tr PA^2 - \tr(PAP)^2\\[2ex]
    &=\tr PA^2-(\tr PA)^2\\[2ex]
    &=\Var_P(A).
    \end{array}
    \]
    An arbitrary state $ \rho $ is by the spectral theorem a convex combination
    $
    \rho=\sum_i \lambda_i P_i
    $
    of pure states. Hence
    \[
    I^c_\rho(A)\le\sum_i \lambda_i I^c_{P_i}(A)=\sum_i \lambda_i \Var_{P_i}(A)\le\Var_\rho(A),
    \]
    where we used the convexity of the metric adjusted skew information and the concavity of the variance.
    \end{proof}

    \subsection{The metric adjusted correlation}

    We have developed the notion of metric adjusted skew information, which is a generalization
    of the Wigner-Yanase-Dyson skew information. It is defined for
    all regular metrics (symmetric and monotone), where the term regular means that the associated
    Morozova-Chentsov functions
    have continuous extensions to the closed first quadrant with finite values everywhere except in the point $ (0,0). $

    \begin{definition}
    Let $ c $ be a regular Morozova-Chentsov function, and let $ d $
    be the representing function (\ref{the function d representing a regular metric}).
    The metric adjusted correlation is defined by
    \[
    \Corr^c_\rho(A, B)=\tr\rho A^*B-{\textstyle\frac{m(c)}{2}}\tr A^*\, d(L_\rho,R_\rho)B
    \]
    for arbitrary matrices $ A $ and $ B. $
    \end{definition}

    Since $ d $ is symmetric, the metric adjusted correlation is
    a symmetric sesqui-linear form which by (\ref{skew information in terms of d}) satisfies
    \[
    \Corr^c_\rho(A, A)=I^c_\rho(A)\qquad\text{for self-adjoint\,} A.
    \]
    The metric adjusted correlation is not a real form on self-adjoint matrices, and it is not
    positive on arbitrary matrices. Therefore, Cauchy-Schwartz inequality only gives a bound
    \begin{equation}\label{correlation inequality}
    |\Re\Corr^c_\rho(A, B)|\le I^c_\rho(A)^{1/2} I^c_\rho(A)^{1/2}
    \le\Var_\rho(A)^{1/2}\cdot \Var_\rho(B)^{1/2}
    \end{equation}
    for the real part of the metric adjusted correlation. However, since
    \[
    \Corr^c_\rho(A, B)-\Corr^c_\rho(B, A)=\tr\rho[A,B]\qquad A^*=A,\, B^*=B,
    \]
    we obtain
    \[
    \frac{1}{2}|\tr\rho[A,B]|=|\Im\Corr^c_\rho(A, B)|
    \]
    for self-adjoint $ A $ and $ B. $ The estimate in (\ref{correlation inequality}) can therefore not be used
    to improve Heisenberg's uncertainty relations\footnote{In the first version of this paper, which appeared
    on July 22 2006, the estimation in (\ref{correlation inequality}) was erroneously extended to the metric
    adjusted skew information itself and not only to the real part, cf. also Luo~\cite{kn:luo:2003}
    and Kosaki~\cite{kn:kosaki:2005}. The author is indebted to Gibilisco and
    Isola for pointing out the mistake.}.

    \subsection{The variant bridge}

    The notion of a regular metric seems to be very important. We note that the Wigner-Yanase-Dyson metrics
    and the Bures metric are regular, while the Kubo metric and the maximal symmetric monotone metric are not.

    The continuously increasing bridge with Morozova-Chentsov functions
    \[
    c_\gamma(x,y)=x^{-\gamma}y^{-\gamma}\left(\frac{x+y}{2}\right)^{2j-1}\qquad 0\le\gamma\le 1
    \]
    connects the Bures metric $ c_0(x,y)=2/(x+y) $ with the maximal symmetric monotone metric $ c_1(x,y)=2xy/(x+y). $
    Since the Bures metric is regular and the maximal symmetric monotone metric is not, any bridge connecting them must
    fail to be regular at some point. However, the above bridge fails to be regular at any point $ \gamma\ne 0. $
    A look at the formula (\ref{canonical representation of c in terms of h}) shows that a symmetric monotone metric
    is regular, if and only if $ \lambda^{-1} $ is integrable with respect to $ h(\lambda)\,d\lambda. $ We may obtain this
    by choosing for example
    \[
    h_p(\lambda)=\left\{\begin{array}{lrl}
                        0,\quad &\lambda &<1-p\\[1ex]
                        p, &\lambda&\ge 1-p
                        \end{array}\right.\qquad 0\le p\le 1
    \]
    instead of the constant weight functions. Since
    \[
    \int\frac{(\lambda^2-1)(1+t^2)}{(1+\lambda^2)(\lambda+t)(1+\lambda t)}\,
    d\lambda=\log\frac{1+\lambda^2}{(\lambda+t)(1+\lambda t)}
    \]
    we are by tedious calculations able to obtain the expression
    \[
    f_p(t)=\frac{1+t}{2}\left(\frac{4(1-p+t)(1+(1-p)t)}{(2-p)^2 (1+t)^2}\right)^p\qquad t>0
    \]
    for the normalized operator monotone functions represented by the $ h_p(\lambda) $ weight functions
    \cite[Theorem 1]{kn:hansen:2006:2}.
    The corresponding Morozova-Chentsov functions are then given by
    \begin{equation}\label{variant increasing bridge}
    c_p(x,y)=\frac{(2-p)^{2p}}{(x+(1-p)y)^p ((1-p)x+y)^p}\left(\frac{x+y}{2}\right)^{2p-1}
    \end{equation}
    for $ 0\le p\le 1. $
    The weight functions $ h_p(\lambda) $ provides a continuously increasing bridge from the zero function
    to the unit function. But we cannot
    be sure that the corresponding Morozova-Chentsov functions are everywhere increasing,
    since we have adjusted the multiplicative constants such that
    all the functions $ f_p(t) $ are normalized to $ f_p(1)=1. $ However, since by calculation
    \[
    \frac{\partial}{\partial p} f_p(t)=
    \frac{-2p^2(1-t)^2}{(2-p)^3(1+t)}
    \left(\frac{4(1-p+t)(1+(1-p)t)}{(2-p)^2 (1+t)^2}\right)^{p-1}<0,
    \]
    we realize that the representing operator monotone functions are decreasing in $ p $ for every $ t>0. $ In conclusion, we have shown that
    the symmetric monotone metrics given by (\ref{variant increasing bridge}) provides a continuously
    increasing bridge between the smallest and largest
    (symmetric and monotone) metrics, and that all the metrics in the bridge are regular except for $ p=1. $

    {\footnotesize



      \vfill

      \noindent Frank Hansen: Department of Economics, University
       of Copenhagen, Studiestraede 6, DK-1455 Copenhagen K, Denmark.}

      \end{document}